\documentclass[12pt]{iopart}
\usepackage[latin1]{inputenc}
\usepackage{graphicx}
\usepackage[T1]{fontenc}
\usepackage{bm}
\usepackage{natbib}
\usepackage{graphics}
\usepackage{subfigure}
\usepackage{float}
\def\newblock{\hskip .11em plus .33em minus .07em}
\newcommand{\mgh}{$\mathrm{MgH^{+}\,}$}

\begin{document}
\title{Rotational cooling of molecules using lamps}
\author{I. S. Vogelius, L. B. Madsen and M. Drewsen\dag}
\address{Department of Physics and Astronomy,
  University of  Aarhus, 8000 {\AA}rhus C, Denmark}
\address{\dag\ Department of Physics and Astronomy,
  University of  Aarhus, 8000 {\AA}rhus C, Denmark and QUANTOP - Danish National Research Foundation Center for
  Quantum Optics}
\date{\today}

\begin{abstract}
We investigate theoretically the application of tailored incoherent far-infrared fields in combination with laser excitation of a single
rovibrational transition for rotational cooling of translationally cold polar diatomic molecules. The cooling schemes are effective on a
timescale shorter than typical unperturbed trapping times in ion traps and comparable to obtainable confinement times of neutral molecules.

\end{abstract}
\pacs{33.80.Ps,33.20.Vq,82.37.Vb}
\maketitle

In recent years, the physics of cold molecules has become a very active research area
 \citep{friedrich,julienne,egorov,meijer3,GrimmMolecularBEC,KetterleMolecularBEC,JinMolecularBEC}.

In this short paper we discuss the possibility to rotationally cool polar molecules by incoherent far-infrared radiation in combination with a
single laser-excited rovibrational transition. Specifically, we consider molecules which are already translationally cold but where the
rotational and vibrational degrees of freedom are in equilibrium with black-body--radiation (BBR) at a temperature of $300$\,K \citep{molhave}.
For lighter molecules, the population will then be distributed over many rotational states, while only the vibrational ground state will be
populated. The presented cooling schemes are, from an experimental point of view, significantly simpler than our previous proposed schemes which
required two laser-induced transitions \citep{vogelius}. The incoherent field is tailored for optimum cooling into the rovibrational ground
state under the constraints set by the spectral density profile of a mercury lamp \citep{kimmitt}. The timescale for the cooling is shorter than
the typical unperturbed trapping time in an ion trap \citep{molhave} and comparable to realistic trapping times for neutral molecules
\citep{meijer1}.

\begin{figure}
  \begin{center}
    \mbox{
      \subfigure[\vspace{-0.2cm}Raman scheme]{\includegraphics[width=0.4\textwidth]{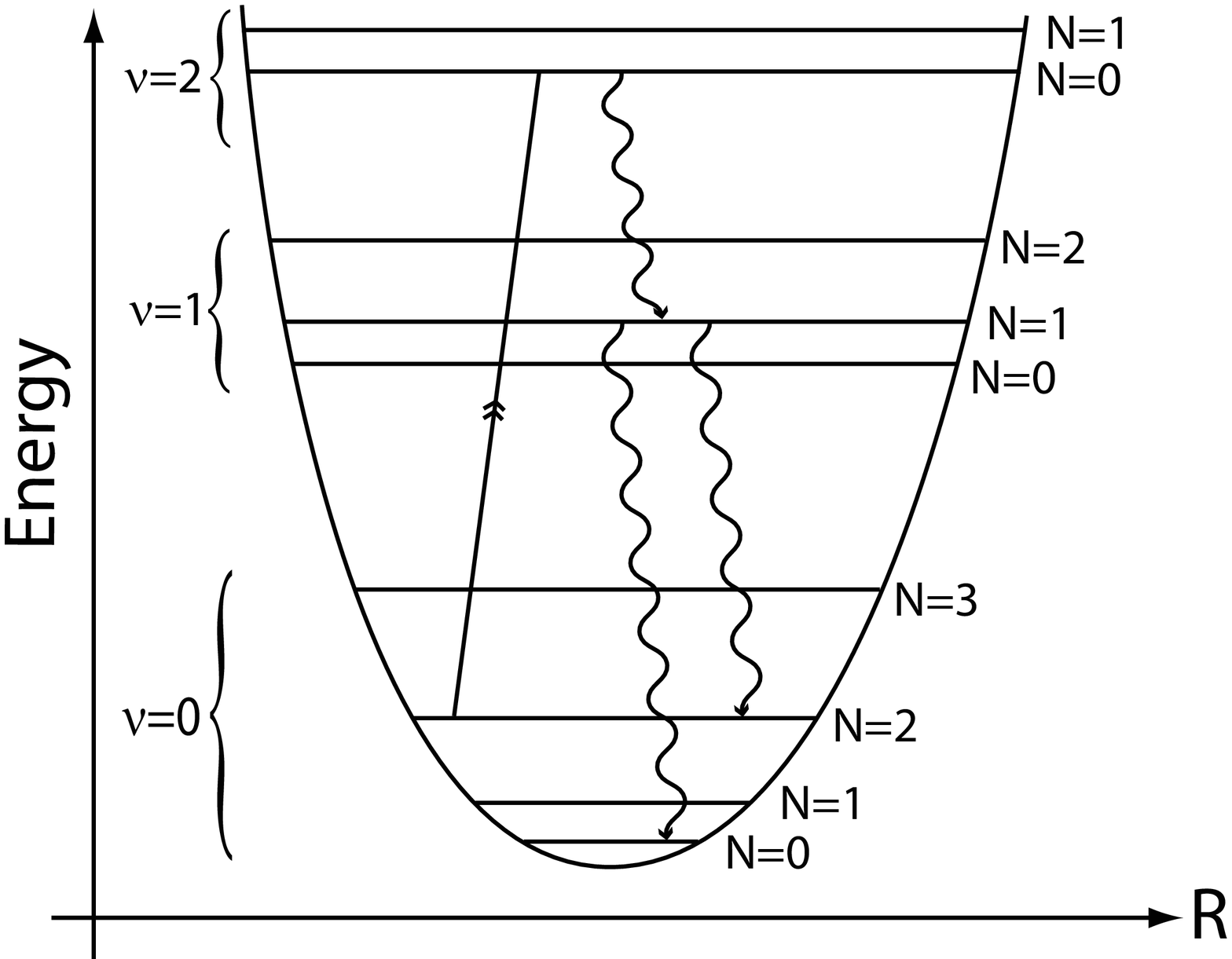}}
    }
    \mbox{
      \subfigure[\vspace{-0.2cm}Direct scheme]{\includegraphics[width=0.4\textwidth]{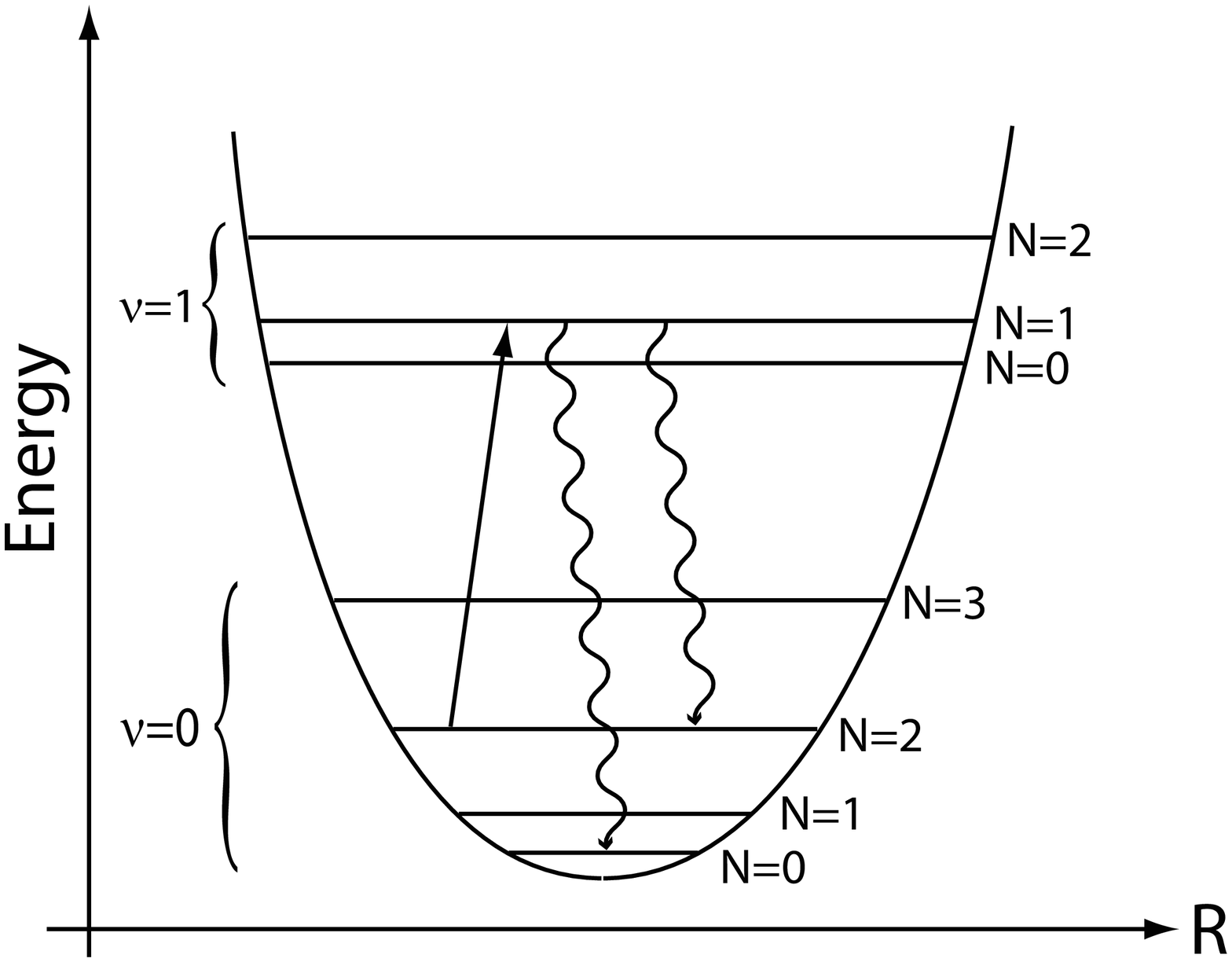}}
    }
    \end{center}

  \caption{Sketch of potential energy curves and rovibrational energy levels showing laser driven transitions and spontaneous decay paths for the (a) Raman
      scheme and (b) direct scheme. Line with double-arrow: transition between
      rovibrational states driven by Raman pulses. Line with single-arrow: direct dipole laser excitations. Wiggly arrows:
      spontaneous decays. The internuclear distance is denoted by $R$.}
\label{fig:schemes}
\end{figure}

Figure \ref{fig:schemes} shows laser excitations and subsequent spontaneous emission paths for our proposed rotational cooling schemes. With
$(\nu,N)$ denoting the vibrational $\nu$ and rotational $N$ quantum numbers, the laser-induced transitions can either be Raman transitions via
an excited electronic state between the levels $(0,2)$ and $(2,0)$ (a) or direct transitions between levels $(0,2)$ and $(1,1)$ (b). In both
cases, the effect of the laser excitations are to pump the molecules from the (0,2) level into the $(0,0)$ ground state through the subsequent
dipole allowed radiative decays. In combination with BBR-induced transitions a significant enhancement in the population of the $(0,0)$ state
can be obtained. At a specific temperature, the strength of the individual BBR-induced transitions are fixed, but by introducing additional
incoherent far-infrared radiation derived from, e.g., a high-pressure mercury lamp, and by applying frequency filters to tailor the radiation
intensity, specific rotational transition rates can be enhanced for improved cooling. One can, for instance, use a high-pass filter
\citep{winnewisser} such that the rate of the $(0,1)\leftrightarrow (0,2)$ cooling transition will be enhanced without effecting the rate of the
$(0,0)\leftrightarrow (0,1)$ heating transition.

Though our schemes are widely applicable, we now focus on implementations in the particular case of \mgh which has been translationally cooled
experimentally \citep{molhave}. We showed previously that the laser intensities needed to saturate the Raman transition depicted in figure
\ref{fig:schemes}(a) are obtained by standard pulsed laser systems  \citep{vogelius}.  For the direct scheme (figure \ref{fig:schemes}(b)), the
infrared field ($\sim
5.9\mu$m for \mgh) could, e.g., be obtained by difference-frequency generation. 
Given the Einstein A-coefficient of $\sim 20$\,s$^{-1}$ for the relevant transition, and assuming a maximum detuning of 1\,GHz due to
fluctuations in the frequency of the laser light, a 1$\mu$J, $10$\,ns pulse focused to a realistic beam spot size of 1\,mm$^2$ will saturate
the transition. 

Turning to the lamp, the presence of this source may improve the cooling rate by speeding up the feeding of the $(0,2)$ pump state from
higher-lying states. The transfer of population away from the pump state is, however, inevitable and it may thus be expected that a certain
spectral density distribution of the incoherent radiation will cool optimally. We now investigate this hypothesis under realistic experimental
constraints. The energy density of light in a standard mercury lamp at the wavelength of interest ($\sim 500 \mu$m, corresponding to low-lying
rotational transitions in \mgh) is similar to a $4000$\,K BBR source \citep{kimmitt}. Assuming unit magnification of the light source and a
collection solid angle of $2\pi$, which seems reasonable using a reflector and a large aperture molded lens
\citep{Nicolaisen_PrivateCommunication}, the intensity of the lamp, in the important part of the spectrum at the position of the molecules,
$I_{lamp}$, can exceed 5 times the intensity of the BBR, $I_{BBR}$, at a temperature of $300$\,K.

As in previous works \citep{vogelius,vogeliusLongPRA}, we model the cooling dynamics by rate equations including transitions induced by the
incoherent radiation as well as by the laser(s). The molecular parameters are obtained by calculating Born-Oppenheimer potential energy curves
and dipole moment functions using standard quantum chemistry codes \citep{gaussian} followed by the calculation of radial wave functions using
the Numerov method \citep{LeroyLevel75}. The nuclear wave function and the electronic dipole moment function determine the Einstein
coefficients. All simulations are made with an initial $300$\,K Boltzmann distribution over molecular energy levels and a pulsed laser system
with a repetition rate of 100\,Hz saturating the driven transition during each pulse.

We maximize the final population in the ground state as function of the incoherent radiation density at the individual rotational transition
frequencies for specific cooling times by the downhill simplex method  \citep{SimplexSearchMethod}.  We find that the density distribution
should be maximal for the $(0,1)\leftrightarrow (0,2)$, $(0,2)\leftrightarrow (0,3)$, and $(0,3) \leftrightarrow (0,4)$ transitions and zero
otherwise, and that the introduction of a time-dependent field only improve the cooling efficiency marginally. This simple shape may be
understood by noting that for a molecule subject to BBR, stimulated processes dominate at low frequencies while spontaneous emission do so at
high frequencies. As a result, the highest populated state, $(\nu,N)_{max,BBR}$ is the one where spontaneous and stimulated transition rates are
balanced ($(\nu,N)_{max,BBR}=(0,4)$ for \mgh). The BBR density is low and transition rates are small for rotational states lying below
$(\nu,N)_{max,BBR}$. Introducing additional incoherent radiation to drive transitions between these low-lying states will accelerate the process
of refilling the $(0,2)$ pump state, and thereby increase the cooling rate. Radiation which couples states above the peak in the $300$\,K BBR
population distribution would heat the distribution as would any radiation in addition to BBR at the $(0,0)\leftrightarrow (0,1)$ transition
frequency.

\begin{figure}[!tp]
\begin{center}
  \includegraphics[width=0.55\textwidth]{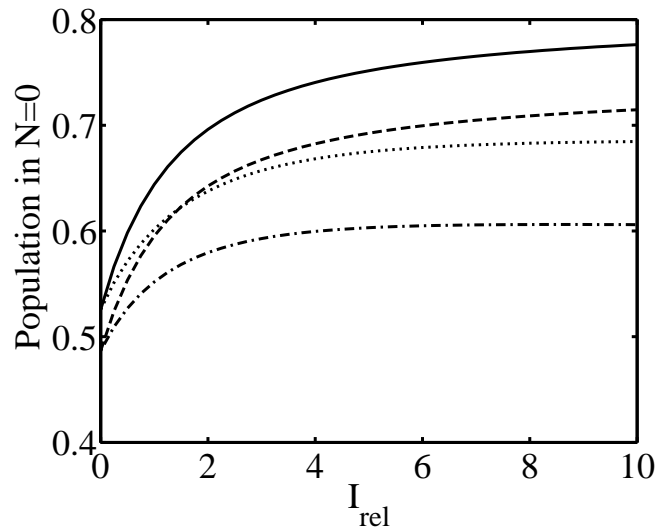}
  \end{center}
  \caption{\label{fig:powerdep} Ground state population of \mgh after
    60 s of cooling vs.\ relative intensity of incoherent
    radiation from a lamp. The schemes are: Direct scheme with optimal incoherent
    radiation distribution (solid) and direct scheme with
    ``quartz-filtered'' distribution (dotted), Raman scheme with
    optimal incoherent radiation distribution (dashed), and Raman
    scheme with ``quartz-filtered'' incoherent radiation distribution
    (dashed-dotted). See text for details.
    Note scale
    of the ordinate.}
\end{figure}

Figure \ref{fig:powerdep} shows the cooling efficiency as a function of $I_{rel}=\frac{I_{lamp}}{I_{BBR}}$. The influence of the incoherent
field on the cooling process is seen to increase significantly up to $I_{rel}\approx 5$ from where only minor improvements may be obtained. This
intensity level coincides well with the estimated obtainable with a mercury lamp.

The schemes discussed here can be extended to $^2\Sigma$, $^3\Sigma$, and $^2\Pi$ electronic ground states without loosing their experimental
feasibility, and should therefore be applicable to almost any diatomic molecule with rotational and vibrational transition rates comparable to
those of \mgh or faster \citep{vogeliusLongPRA}.

In conclusion, we have demonstrated that polar molecules trapped in a room temperature environment can be cooled rotationally by the combination
of a lamp inducing selected rotational transitions and a laser driving a single rovibrational transition. Under realistic constraints we have
shown that there exists a specific tailored incoherent frequency distribution which optimizes the cooling process. From an experimental
view-point, the considered schemes are very simple, and hence attractive for sympathetically cooled target molecular ions, and potentially also
for trapped neutral molecules and molecular ions in storage rings.

L. B. M. is supported by the Danish Natural Science Research Council (grant no 21-03-0163). M.D. acknowledges financial support from the Danish
National Research Foundation through the Quantum Optics Center QUANTOP.

\bibliographystyle{jphysicsb}

\end{document}